\documentclass[aps,pre,groupedaddress,showpacs,twocolumn]{revtex4}

\usepackage{amsmath}

\usepackage{amssymb}

\usepackage{amsbsy}

\usepackage{graphicx}

\usepackage{color}

\usepackage{hyperref}

\hypersetup{
bookmarks=false,
pdftitle={Nonlinear compactons in sawtooth lattices},
anchorcolor=blue,
colorlinks=true,
citecolor=blue,
urlcolor=blue,
linkcolor=blue}

\begin{document}

\title{Compactification tuning for nonlinear localized modes in sawtooth 
lattices}

\author{Magnus Johansson}\email{mjn@ifm.liu.se}
\homepage{https://people.ifm.liu.se/majoh}
\affiliation{Department of Physics, Chemistry and Biology (IFM), Link\"{o}ping 
University, SE-581 83 Link\"{o}ping, Sweden}

\author{Uta Naether}\email{naether@unizar.es}
\affiliation{Instituto de Ciencia de Materiales de Arag\'on y
 Departamento de F\'{\i}sica de la Materia Condensada,
 CSIC-Universidad de Zaragoza, Zaragoza, E-50009, Spain.}

\author{Rodrigo A. Vicencio}\email{rodrigov@uchile.cl}
\affiliation{Departamento de F\'isica and MSI-Nucleus on Advanced Optics, Center for Optics and Photonics (CEFOP), Facultad de Ciencias, Universidad de Chile, Santiago, 7800003, Chile}

\begin{abstract}
We discuss the properties of nonlinear localized modes in sawtooth lattices, 
in the framework of a discrete nonlinear Schr{\"o}dinger model with general 
on-site nonlinearity. Analytic conditions for existence of exact compact 
three-site solutions are obtained, and explicitly illustrated for the cases of 
power-law (cubic) and saturable nonlinearities. 
These nonlinear compact modes appear as continuations of linear compact modes 
belonging to a flat dispersion band. While for the linear system a 
compact mode exists only for one specific ratio of the two different coupling 
constants, nonlinearity may lead to compactification of otherwise non-compact 
localized modes for a range of coupling ratios, at some specific power. For 
saturable lattices, the compactification power can be tuned by also varying 
the nonlinear parameter. Introducing different on-site energies and anisotropic 
couplings yield further possibilities for compactness tuning. The properties 
of strongly localized modes are investigated numerically for cubic and 
saturable nonlinearities, and in particular their stability over large 
parameter regimes is shown. Since the linear flat band is isolated, 
its compact modes may be continued into compact 
nonlinear modes both for focusing and defocusing nonlinearities. Results are 
discussed in relation to recent realizations of sawtooth photonic lattices. 
\end{abstract}

\date{\today}
\pacs{63.20.Pw,  42.82.Et, 78.67.Pt, 03.75.Lm} 
\maketitle

\section{Introduction}
\label{sec:introduction}

The creation and manipulation of nonlinear localized lattice excitations 
(``discrete solitons'' or ``discrete breathers'') is of
interest in many areas of physics \cite{FG08}, and in particular within 
nonlinear optics \cite{Lederer08}.  In many 
contexts, one strives for achieving optimal spatial localization. 
In the presence of linear dispersion (diffraction), resulting, e.g., from a 
linear (harmonic) coupling between neighboring sites, the localization will 
generically be exponential. However, non-generic situations may appear 
in presence of nonlinear dispersion, e.g., when the inter-site interaction 
also has a nonlinear contribution. In such situations, a proper control 
of the nonlinear effects may lead to a complete cancellation of the linear 
coupling between sites, resulting in exact ``discrete compactons'' fully
localized at a small number of sites, with strictly zero tail (for examples, 
see, e.g., \cite{KK02,OJE03,AKS10}). 
The experimental realization of these 
specific nonlinear interactions poses however serious challenges, and so far 
no direct observation of compactification of nonlinear lattice modes 
has, to the best of our knowledge,  been reported. 

On the other hand, it is well known that in certain lattices geometries, 
frustration effects make possible the existence of strictly compact modes 
even in absence of any nonlinearity. Such modes can be seen to result from
destructive interference of couplings between different sites, and are 
typically associated with a flat linear dispersion band. A number of 
different examples of this scenario are discussed, e.g., in 
Refs.\ \cite{Derzhko10,HA10,Flach14,Liu14}, and references therein. In 
particular, recent experiments have being performed in two-dimensional (2D) 
Lieb photonic lattices, where different transport regimes~\cite{lieb1} as well 
as the excitation of the localized flat band mode~\cite{lieb2,lieb3} 
were observed. 
Maybe the most well known example is the 2D Kagome lattice, which  supports 
modes strictly localized on six sites along a hexagonal geometry 
(see, e.g., \cite{BWB08}). As was shown in 
Ref.\ \cite{VJ13}, these compact modes exist also in presence of a pure on-site 
nonlinearity, and indeed constitute an effective ground state of the system 
in the small-power regime for a defocusing Kerr (cubic) nonlinearity. However, 
above a rather small threshold power the compact modes destabilize, and 
transform into stable standard localized modes with exponential tails.

The compactification of linear modes due to competing linear interactions may 
appear also in one-dimensional (1D) settings, and a number of examples have 
been illustrated, e.g., in Refs.\ \cite{Derzhko10,Flach14}. One of the 
most studied 1D lattices is the sawtooth chain; see, e.g., 
\cite{SSWC96,SHSRC02,ZT04,HA10,Derzhko10} (also called $\Delta$ chain 
\cite{NK96}
or triangle lattice \cite{HAM13,Flach14}). Due to its construction from 
corner-sharing triangles, it is often considered as a 1D analogue of the 
Kagome lattice, and is one of the simplest possible lattices allowing 
for compact linear modes and a flat band, which makes it 
particularly attracting for experimental realizations. A compact mode in a 
similar setting was used to experimentally observe a Fano 
resonance~\cite{Weimann13}. Very recently, 
the linear properties of a sawtooth photonic lattice, created using the 
femtosecond laser written technique, were also studied 
experimentally \cite{photonics14}. 
Furthermore, the dependency of coupling  constants of these lattices on 
polarization~\cite{Rojas14} provides ample and tunable opportunities beyond 
geometric considerations. This is important since, in contrast to the Kagome 
lattice, compactification in a linear 
sawtooth lattice appears only at one specific (irrational) ratio between 
the two different coupling constants (horizontal and diagonal), which again 
poses an experimental challenge for fine-tuning. However, as we will show 
in this work, nonlinearity allows the existence of compact modes for different 
coupling ratios, which certainly would facilitate the observation of highly 
localized states in a real experiment.

The aim of the present paper is to illustrate how nonlinearities and other 
effects can be used for the purpose of facilitating the compactification 
tuning of localized modes in sawtooth lattices. In Sec.\ \ref{sec:general} 
we describe the general model, using the discrete  nonlinear Schr{\"o}dinger 
(DNLS) equation with a general on-site nonlinearity and linear 
coupling constants representing the sawtooth geometry. We discuss the 
linear dispersion relation and its flat-band structure, and obtain 
the general conditions on the geometry for existence of 3-site compactons as 
exact stationary 
solutions. In Sec.\ \ref{sec:power-law} we specialize to power-law 
nonlinearities, and obtain an expression for the power needed for 
compactification, valid for a whole range of ratios for the linear coupling 
constants. For the physically most interesting case of cubic Kerr-nonlinearity 
(focusing or defocusing), 
we perform a detailed numerical analysis of localized modes (not 
restricted to compactons) in Sec.\ \ref{sec:cubic}, illustrating explicitly 
how certain, linearly stable, localized modes compactify when the appropriate 
relation between coupling ratios and power is fulfilled. Another 
physically important case of a saturable nonlinearity is analyzed in 
Sec.\ \ref{sec:saturable}. The saturability introduces another parameter 
which can be employed, together with the power, for compactification tuning, 
and we illustrate the scenarios in different regimes analytically as 
well as numerically, particularly showing the existence of strongly 
localized, stable modes over large parameter regimes. 
Finally, Sec.\ \ref{sec:further} discusses two additional generalizations
of the sawtooth lattice with cubic nonlinearity, yielding further 
possibilities for compactification tuning: on-site energy alternations 
(Sec.\ \ref{sec:onsite}) and coupling anisotropies 
(Sec.\ \ref{sec:anisotropic}). 
Conclusions are given in 
Sec.\ \ref{sec:conclusions}. 

\section{Model and general compacton conditions}
\label{sec:general}

\begin{figure}
\includegraphics[width=0.5\textwidth]{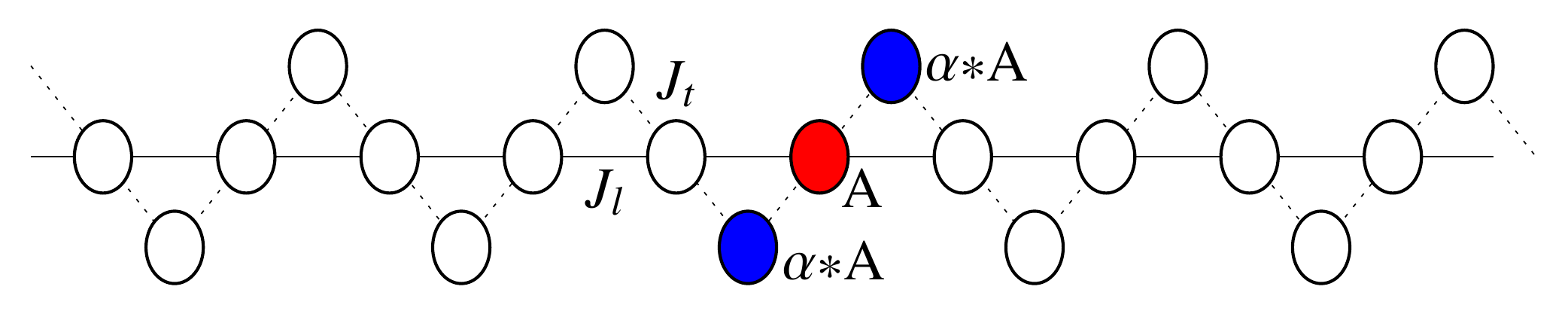}
\caption{(Color online) Geometry of the sawtooth lattice with its compact mode 
(white circles imply zero amplitudes).}
\label{fig:sawtooth}
\end{figure}

We consider the following form of the discrete nonlinear Schr\"odinger equation on a sawtooth lattice sketched in Fig.~\ref{fig:sawtooth},

\begin{equation}
i \dot{u}_n + \sum_{m\neq n} J_{n,m} u_m + \gamma f(|u_n|^2) u_n=0\ ,
\label{dnls}
\end{equation}
where $n=1,2,\dots, N$. The coupling $J_{n,m}$ denotes the interactions between the site $n$ and its connected lattice sites $m$, with diagonal (``tip'') couplings $J_t$ and horizontal (``line'') couplings 
$J_l$. The function $f(x)$, describing a general on-site nonlinearity with 
strength $\gamma$, is at this point arbitrary. 
We may put $f(0)=0$, since any homogeneous linear on-site potential may be 
removed by a gauge transformation (extensions to non-homogeneous on-site 
potentials will be discussed in 
Sec.~\ref{sec:further}). Generally, the parameters $J_l$, $J_t$ and $\gamma$ 
may be complex-valued (describing effects of gain and/or loss), 
but in our explicit examples in  
Secs.~\ref{sec:power-law}-\ref{sec:further} we will only consider real-valued  
cases. We should remark that the geometry in Fig.~\ref{fig:sawtooth} 
apparently differs from that usually associated with sawtooth 
lattices~\cite{NK96,SSWC96}. Instead of taking equally oriented triangles, we 
take the triangle tips to point alternately up and down from the horizontally 
line-coupled sites. In this way, we avoid the otherwise physically unavoidable 
interaction between neighboring vertices, describing a more appropriate 
experimental realization of a photonic sawtooth lattice~\cite{photonics14}. 
This obviously does not change the mathematical model, which in both cases 
only contains the two nearest-neighbour couplings $J_t$ and $J_l$.  

Looking for 
stationary solutions
of the form
%
\begin{equation}
u_n(t) = a_n e^{i \lambda t}\ , 
\label{eq:stat}
\end{equation}
%
the first term in Eq.\ (\ref{dnls}) is replaced with $-\lambda u_n$. 
Assuming the line sites to be of odd $n$, whereas for the tips $n$ is even,  
then leads to
the following set of equations in the linear regime ($\gamma |a_n|^2= 0$):
%
\begin{equation*}
\lambda a_n=J_t(a_{n+1}+a_{n-1})+\begin{cases} J_l (a_{n+2}+a_{n-2})& \text{ for $n$ 
odd\ ,}\\ 0& \text{ for $n$ even\ .} \end{cases}
\end{equation*}

The linear dispersion relation \cite{Derzhko10,HA10,photonics14} is computed 
by assuming a propagating plane wave $a_n=A {\rm e}^{ikn}$, obtaining for the 
linear frequencies, rescaled with the horizontal coupling constant $J_l$, 
%
\begin{eqnarray}
\bar{\lambda}_\pm
(k; J) &\equiv& \lambda_{\pm}/J_l =
\label{eq:disp}\\ &&\cos(2k) \pm \sqrt{\cos ^2(2k)+2 J^2 \left[1+\cos (2k)\right]}\ .
\nonumber
\end{eqnarray}
Here we have also introduced the rescaled, dimensionless coupling ratio 
$J \equiv J_t/J_l$, which is the essential  system parameter that determines 
its 
qualitative physical properties in the linear regime.

\begin{figure}
\includegraphics[width=0.5\textwidth]{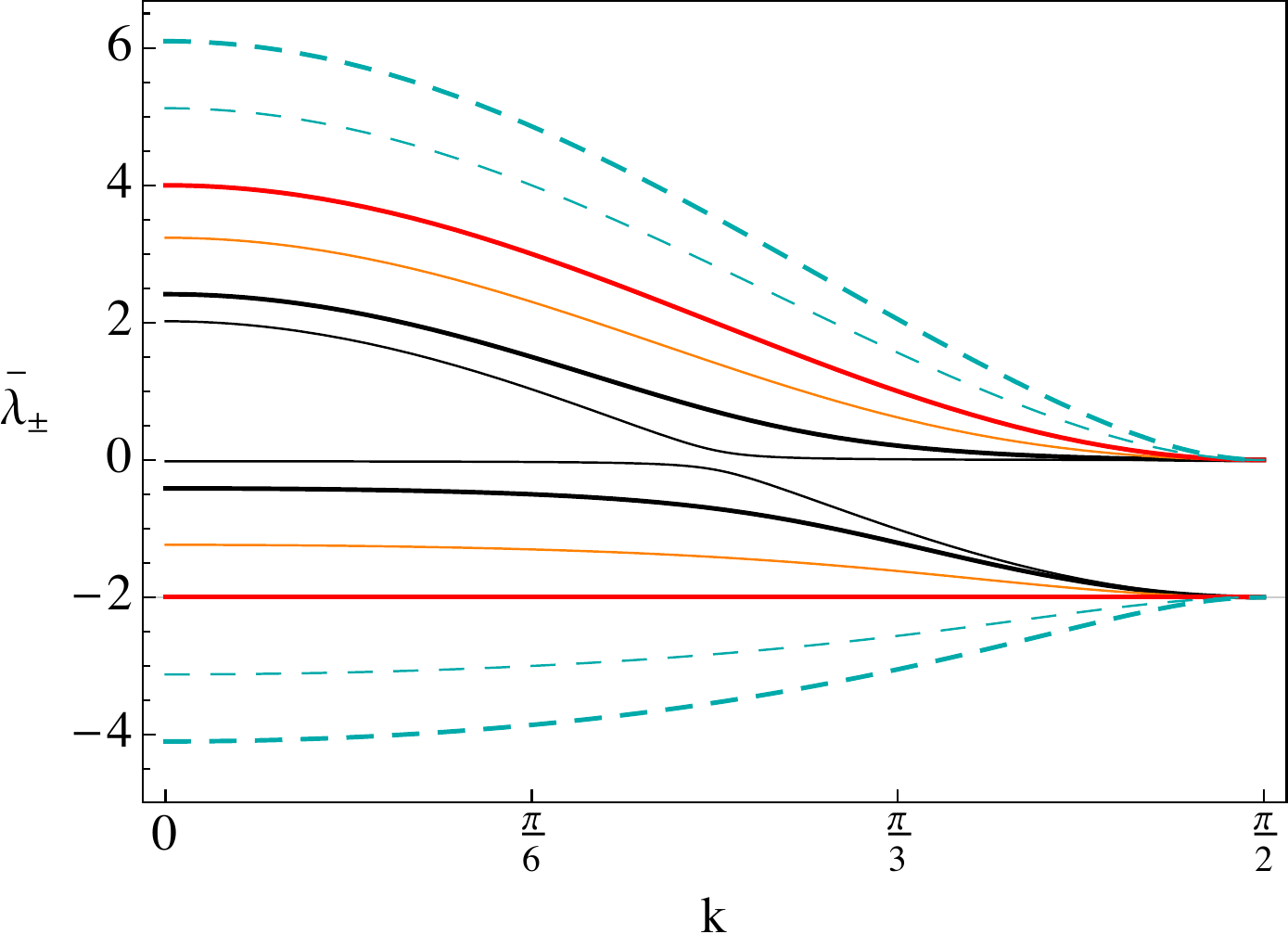}
\caption{(Color online) Dispersion relation (\ref{eq:disp}) for the 
sawtooth lattice for different coupling ratios:
$J$= 0.1 (0.5, 1, $\sqrt{2}$, 2, 2.5) in thin black (thick black, 
thin orange, thick red,  thin dashed, thick dashed).  
}
\label{fig:dispersion}
\end{figure}
%
As illustrated in  Fig.\ \ref{fig:dispersion}, the lower band becomes flat at 
the particular coupling ratio $J = \sqrt{2}$ for which $\bar{\lambda}_- = -2$. 
It is important to note that the flat band existing at these parameter values 
is isolated from the upper band ($\bar{\lambda}_+ = 4 \cos^2 k$) with a 
bandgap of $2$. Thus, in contrast to the Kagome lattice studied in 
Ref.~\cite{VJ13}, where the flat band is always attached to the edge of a 
dispersive band, we may here expect a resonance-free continuation of compact 
modes belonging to the flat linear band for focusing as well as for defocusing 
nonlinearities.

To obtain a condition for existence of compact stationary solutions 
(\ref{eq:stat})
of Eq.\ (\ref{dnls}) for general nonlinearities $f(x)$, we write an ansatz for the solution as a 3-site compacton (analogously to the linear flat-band modes~\cite{Derzhko10,HA10,photonics14}), with $a_{n_0} = A$ for the central, 4-bonded, ``line site'' (red disk in Fig.\ \ref{fig:sawtooth}), and $a_{n_0\pm 1} = \alpha A$ for the two neighboring, 2-bonded, ``tip sites'' (blue in Fig.~\ref{fig:sawtooth}). The condition to have zero amplitude on the neighboring line sites 
$ n_0\pm 2$, and thus also for the rest of the lattice,
then gives immediately an expression for the amplitude ratio $\alpha$ in terms 
of the ratio between the coupling constants: 
\begin{equation}
\alpha = - 1/J\ .
\label{eq:alpha}
\end{equation}
For the central line site $ n_0 $, Eqs.\ (\ref{dnls})-(\ref{eq:stat}) give 
the following expression for the rescaled frequency $\lambda/J_l$:
\begin{equation}
\lambda/J_l = -2  + \frac{\gamma}{J_l} f(|A|^2)\ .
\label{n0}
\end{equation}
The corresponding equations for the two tip sites  $ n_0\pm 1$ yield
\begin{equation}
\lambda/J_l = - J^2 + \frac{\gamma}{J_l}f(|A/J|^2)\ .
\label{n0pm1}
\end{equation}
Eliminating $\lambda$ from Eqs.~(\ref{n0})-(\ref{n0pm1}) gives an implicit, 
general 
condition to determine the central-site intensity $|A|^2$ in terms 
of the coupling ratio $J$ and the ratio $\gamma/J_l$ between nonlinearity 
strength and horizontal coupling constant: 
\begin{equation}
\frac{\gamma}{J_l} \left( f(|A|^2) - f(|A/J|^2) \right) = 2-J^2\ .
\label{gencond}
\end{equation}
Obviously, in the linear limit $\gamma |A|^2 \rightarrow 0$ and the left-hand 
side of Eq.\ (\ref{gencond}) vanishes. This implies that $J_t \rightarrow 
\pm \sqrt{2} J_l$, recovering the linear flat-band condition. Using Eq.\ (\ref{eq:alpha}), we may also determine the total power, $P \equiv\sum_n |u_n|^2$, of the compacton as
\begin{equation}
P = \left(1+2|1/J|^2\right) |A|^2\ .
\label{power}
\end{equation}
Apart form the power, the other conserved quantity of DNLS-like systems is the Hamiltonian
\begin{equation}
H =
\sum_{n\neq m}J_{n,m}u_n^*u_m +\gamma\sum_n F(|u_n|^2)
\label{hamiltonian}
\end{equation}
for the canonical variables $(u_n,-iu_n^*)$. 
The function $F(|u_n|^2)$ defines the nonlinearity of the system 
($f(x)=F'(x)$) and will be 
specified later on. (Note that we chose here a sign-convention in accordance 
with the sign-convention used in the definition of the frequency $\lambda$ in 
(\ref{eq:stat}). As illustrated below, stable compactons may then appear
as ground states for defocusing nonlinearities.)

In order to compare how compactons are related to eventual \textit{ordinary} 
localized stationary modes, which have nonvanishing tails, we implement a 
Newton-Raphson iterative method to find numerical solutions at a given 
accuracy. We will look for fundamental solutions centered at tip or line sites 
or in between sites, using periodic boundary conditions to avoid surface 
effects. A typical lattice size used numerically is $N=40$, while also 
checking the consistency of the results by varying the chain length. We 
generally use $J_l=1$ in all numerical calculations.

For any nonlinear stationary mode $\lbrace a_n\rbrace$ fulfilling 
\eqref{dnls}-\eqref{eq:stat}, 
we perform a standard linear stability analysis~\cite{sta}: we use the 
ansatz $u_n(t) = [a_n+\epsilon_n(t)] e^{i \lambda t}$, and linearize equations in 
$\epsilon_n(t)$, giving access to the spectrum $\lbrace \omega_{\rm l}\rbrace$ 
of perturbations ${\rm e}^{i\omega_{\rm l} t}$. We define a stability parameter 
$g\equiv {\rm Max|Im(\omega_l)|}$, which will be nonzero for unstable modes, 
whereas $g=0$ for stable solutions. To compare the localization properties of 
different modes, we use the participation number defined as
\begin{equation*}
R\equiv\frac{P^2}{\sum_n^N |u_n|^4}\rightarrow \begin{cases} 1 & \text{for a single excited site,}\\
N & \text{for extended homogeneous profiles.}
\end{cases}
\end{equation*}
This parameter gives a measure of the effective size of a given profile. 
In particular, for the compacton solution this formula reduces to
\begin{equation}
R_{c}=\frac{\left[\left|J\right|^2+2\right]^2}{\left|J\right|^4+2}\rightarrow\begin{cases} 
2 & \text{ for } J \rightarrow 0\ ,\\
1 & \text{ for } J\rightarrow \infty\ .
\end{cases}
\label{R2}
\end{equation}
From this, it also follows that the maximum compacton participation number 
$R_c=3$ would appear for $J \rightarrow 1$, 
when the profile would be equally distributed 
between all 3 sites. For the linear case, $R_c(|J|=\sqrt{2})=8/3$.

\section{Power-law nonlinearities}
\label{sec:power-law}

Specializing first to a general, pure power-law nonlinearity, we have 
$f(x) = x^\sigma$. Plugging this into Eq.\ (\ref{gencond}) yields that an exact 3-site compacton exists when the following relation, between the central-site intensity and the coupling constants, is fulfilled:
\begin{equation}
\frac{\gamma |A|^{2 \sigma}}{J_l} =\frac{2  -J^2}{1-1/|J|^{2\sigma}}\ .
\end{equation}
Using Eq.\ (\ref{power}), we find that the compacton condition can alternatively be written in terms of the total power as:
\begin{equation}
P = \left( 2+|J|^2  \right) 
\left(\frac{J_l(2 -J^2)}{\gamma \left(|J|^{2\sigma}-1\right)}\right)^{1/\sigma}.
\label{eq:Psigma}
\end{equation}
For power-law nonlinearities, $F(x)=x^{\sigma+1}/(\sigma+1)$ in Hamiltonian 
\eqref{hamiltonian}; for the compact mode we find 
\begin{eqnarray}
&&H_c\equiv\frac{H}{J_l}=\label{Hc}\\&&- \frac{4 J P}{2 + |J|^2} + \frac{\gamma/J_l}{\sigma+1} 
\left(1 + \frac{2}{ (-J)^{\sigma+1}}\right) \left(\frac{|J|^2 P}
{ 2 + |J|^2}\right)^{(\sigma+1)/2}.\nonumber
\end{eqnarray}

\subsection{Localized modes for cubic nonlinearities}
\label{sec:cubic}

For the special case $\sigma=1$, which corresponds to a standard cubic 
(Kerr) nonlinearity, and real-valued coupling constants, 
Eq.~(\ref{eq:Psigma}) considerably simplifies. 
We can define an effective nonlinearity parameter as
\begin{equation}
\Gamma \equiv \frac{P \gamma}{J_l}= \frac{4-J^4}{J^2 - 1}\ .
\label{Pcubic} 
\end{equation}
%
For Kerr nonlinearities, $\Gamma$ is the only free parameter.
Therefore, for focusing nonlinearities $\Gamma>0$, whereas  $\Gamma<0$ for the defocusing case. We can also calculate the rescaled 
frequency of the (nonlinear) compacton, given by 
\begin{equation}
\bar{\lambda}_c\equiv \frac{\lambda_c}{J_l}=\frac{J^4-2}{1-J^2}.
\label{lamcubic} 
\end{equation}
%
\begin{figure}[t]
\includegraphics[width=0.5\textwidth]{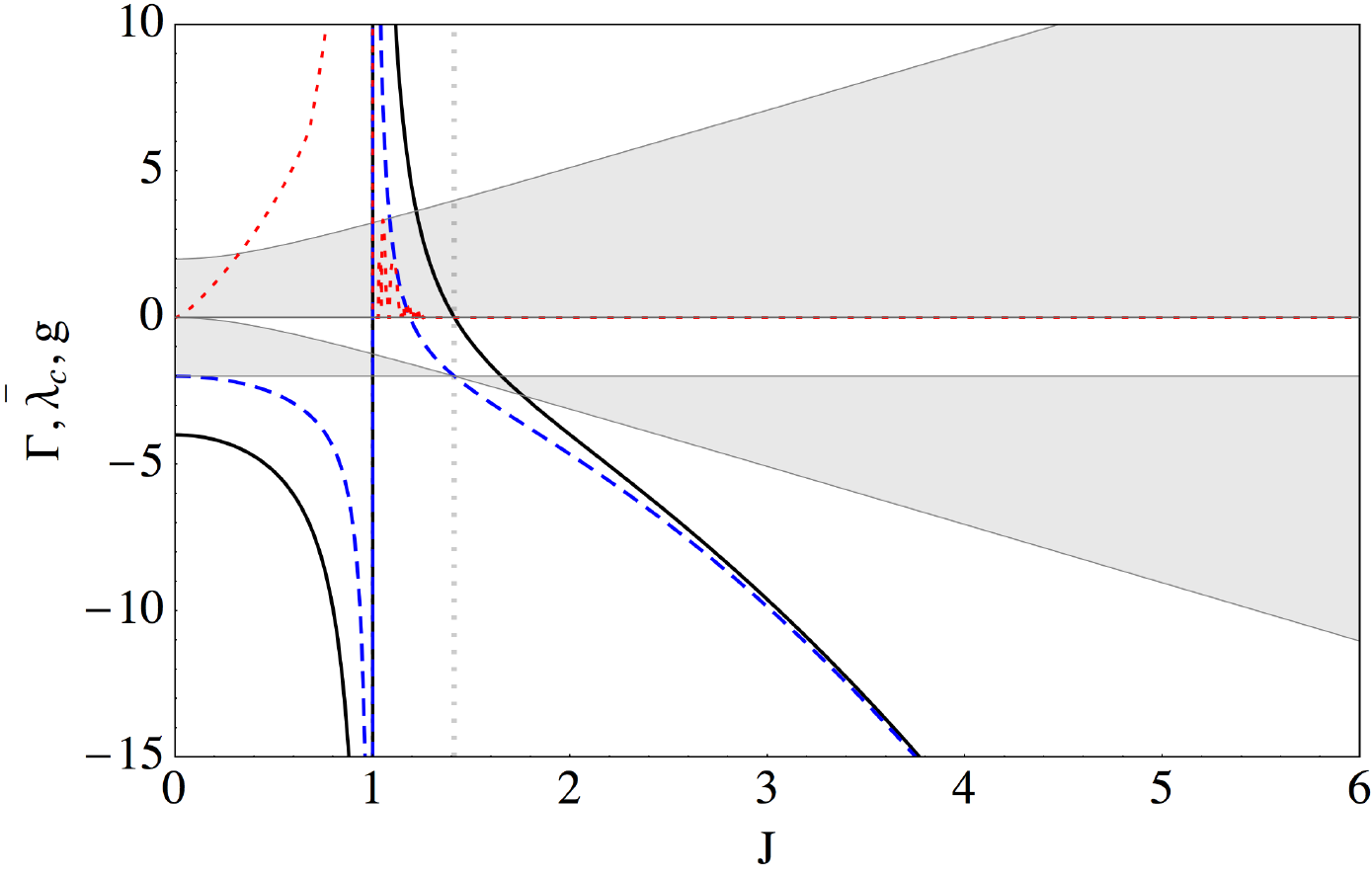}
\caption{(Color online) Properties of the compacton: effective nonlinearity $\Gamma$ (solid), rescaled frequency $\bar{\lambda}_c$ (dashed) and stability parameter $g$ (dotted) 
vs.\ $J$. The linear band extension is shown with shaded areas. The analytical 
expressions for $\Gamma$ and $\bar{\lambda}_c$ are equivalent to those 
of a three-core coupler studied in \cite{SMTL92} for different regimes:
$0<J<1 \leftrightarrow \textrm{``branch c''}$; 
$1<J<\sqrt{2} \leftrightarrow \textrm{``branch b''}$;
$J>\sqrt{2} \leftrightarrow \textrm{``branch d''}$. 
}
\label{fig:kerrcomp}
\end{figure}
The behavior of both $\Gamma$ and $\bar{\lambda}_c$  is shown in 
Fig.~\ref{fig:kerrcomp} with a solid black and dashed line, respectively. 
Furthermore, we show the stability parameter $g$ (red dotted) vs.\  $J$. 
Whenever $g>0$, the solution is unstable. The linear band extensions are 
plotted as shaded background. 
When the linear compacton condition $J = \sqrt{2}$ is fulfilled, 
obviously $\Gamma(J)=0$ and $\bar{\lambda}_c=-2$. At this value, we observe that the 
lower band width becomes strictly zero, corresponding to the perfectly flat 
band of this lattice, occurring for this special geometric relation. 

For $J>\sqrt{2}$, the linear band broadens but a compact 
mode still exists for negative $\Gamma$, with a frequency below the lower band. 
As is seen, the mode stays linearly stable, which was also confirmed by direct 
numerical integration of Eq.\ (\ref{dnls}). 
In the large-$J$ limit, $|\alpha|$ decreases and the compacton, having its 
dominating amplitude on a line site, approaches a 
highly nonlinear single-site localized mode of an ordinary defocusing 
1D chain (as the line-couplings become negligible next-nearest neighbor 
couplings), which is always stable. 

On the other hand, for decreasing $J<\sqrt{2}$ compactons instead occur for a focusing 
nonlinearity ($\Gamma > 0$) and with the frequency 
entering into the gap above the lower band. As can be seen, the mode initially 
remains linearly stable but becomes unstable in a bifurcation at 
$ J\simeq 1.27$.
The origin of this instability can be understood by comparing with the 
well-known properties of the stationary solutions of a three-core coupler 
with focusing nonlinearity (with open boundary conditions, i.e., without any 
surrounding lattice),  studied, e.g., in Ref.\ \cite{SMTL92}. The analytical 
forms of the compactons in the sawtooth lattice are  mathematically equivalent 
to those of 
the symmetric three-core eigenmodes given in \cite {SMTL92}; however, the 
stability properties may differ due to additional possible resonances with 
the surrounding lattice.  As the corresponding modes 
in the focusing three-core coupler 
(``branch b'' with the notation of \cite{SMTL92})  are always stable, the 
observed instabilities for the compacton in the focusing sawtooth lattice 
must originate in resonances between internal oscillations of the compacton 
core and linear oscillations of the surrounding chain. Decreasing $J$ further
towards the singularity at $J=1$, 
stronger instabilities appear as the compacton frequency becomes positive 
and enters the upper linear band. Compactons with positive 
frequency, in a strongly nonlinear focusing lattice, are found to be 
always unstable.

From the above, we reach the important conclusion that there are 
{\em stable compact nonlinear modes in direct continuation of the linear 
flat-band modes for both focusing and defocusing nonlinearities}.  
At the singularity $J=1$ both $\Gamma$ and $\bar{\lambda}_c$ change sign,  and we 
see that for $J<1$, compact modes with negative 
effective nonlinearity and frequency are always unstable. 
Noting that the anti-phased compact modes in this defocusing regime are 
mathematically equivalent to three-core in-phase modes with focusing 
nonlinearity, these instabilities are equivalent to those described 
in Ref.\ \cite{SMTL92} for the solution ``branch c'' 
(corresponding to main amplitudes at the two tip sites). Thus, this
instability originates from an internal resonance between oscillation 
modes of the three-site compacton core.
So to conclude this part, we see that for {\em any} sawtooth 
lattice geometry we can find a compact mode centered on a line site. 
This mode is stable for a wide window in parameter 
space but destabilizes when the surrounding tip site amplitudes become 
comparable to, or larger than, the amplitude of the central line site. 

\begin{figure}[t]
\includegraphics[width=0.5\textwidth]{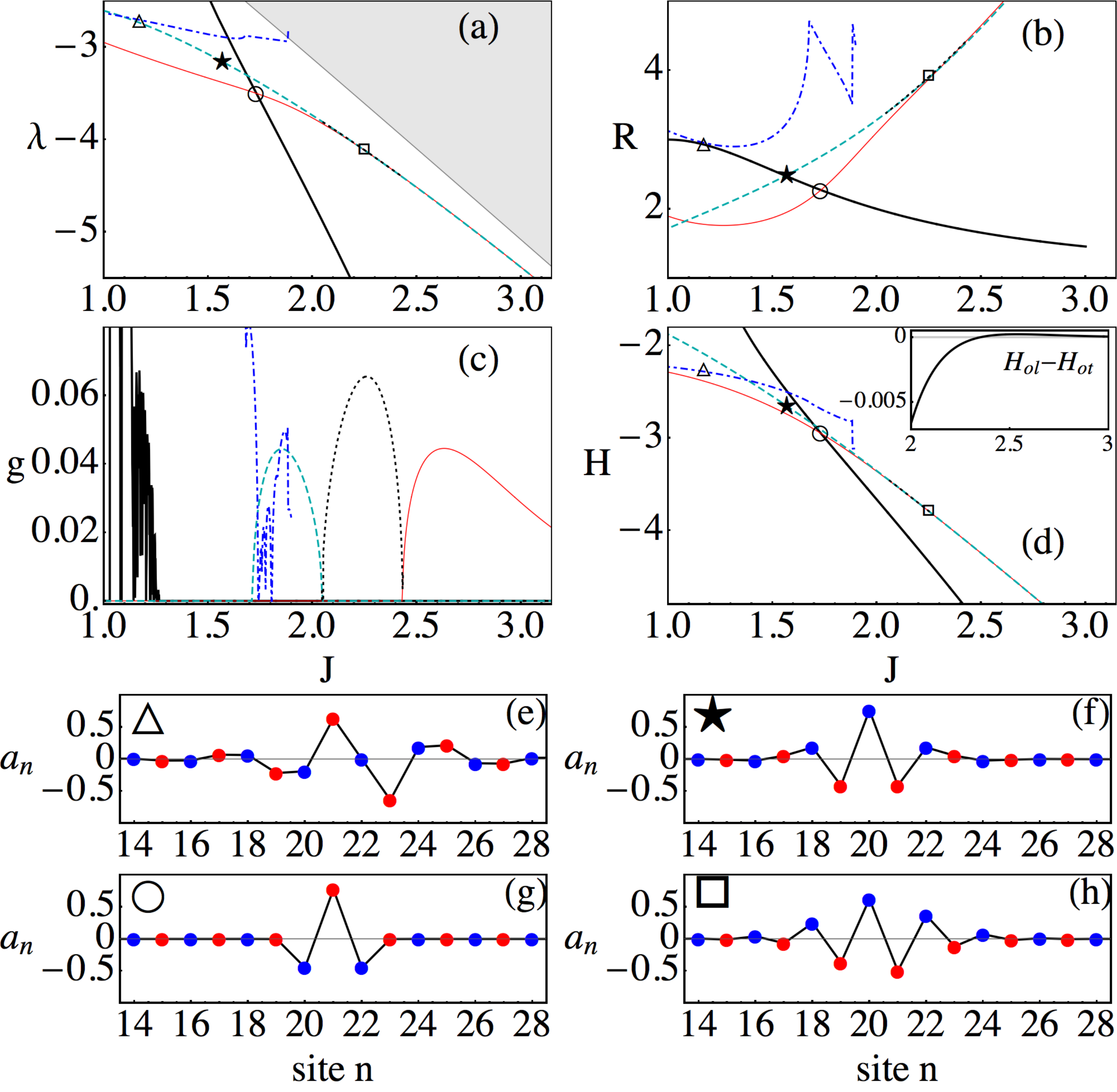}
\caption{(Color online) Properties of defocusing nonlinear modes: 
(a) frequency $\lambda$, (b) $R$, (c) stability parameter $g$  and (d) $H$ vs.\ $J$ for the 
compact, odd line, odd tip, even line and intermediate modes in thick, thin, dashed, 
dotdashed and dotted lines, respectively. The linear band in (a) is shown with a shaded area. In the inset of (d) we show the difference between $H$ for the odd line and tip mode.
For all non-compact modes $\Gamma=-2.5$. The symbols represent the parameter values of the exemplary modes shown below: (e)  $a_n$ of the even line mode, (f) $a_n$ of the odd tip mode, (g) $a_n$ of the odd line mode, (h) $a_n$ of the intermediate mode. Red and blue filled circles represent the amplitude at line and tip sites, respectively.
}
\label{fig:kerrloc}
\end{figure}

%
The nonlinear compactons described above are not isolated solutions for 
a chain with a given $J$, but rather 
belong to certain families of, generally non-compact, nonlinear localized 
modes which may be tuned to compactness by varying the effective nonlinearity 
(e.g., by power tuning) so to fulfill the compacton condition (\ref{Pcubic}). 
We first illustrate this scenario for an effectively {\em defocusing} 
nonlinearity, 
$\Gamma < 0$, in Fig.~\ref{fig:kerrloc}. 
The relations of  $\lambda$,  $R$, $g$ and $H$ vs.\ $J$, for four families of
ordinary localized 
modes with  $\Gamma=-2.5$, in comparison to the compact mode with its 
respective $\Gamma$ given by (\ref{Pcubic}), are shown in 
Fig.~\ref{fig:kerrloc} (a)-(d). Here, the  thick (thin, dashed, dotdashed) 
line represents the compact (odd line, odd tip and even line) mode; some 
examples for these modes are shown in Figs.~\ref{fig:kerrloc} (e)-(g), with 
parameter values corresponding to the respective symbols. 
The odd line mode is centered at a line site and has the same symmetry as the 
compacton, while the even line mode, as well as the odd tip mode, have 
different symmetries being centered around or at a tip site (the even line mode has its 
main amplitudes at the surrounding line sites, while the odd tip mode is 
primarily localized at the central tip site).  
As usual in cubic lattices, the even line mode is unstable everywhere. 
An example for its profile $a_n$ is shown in Fig.\ \ref{fig:kerrloc} (e).
We chose parameter values, where its participation number $R$ 
is very close to the compact mode, but, there is a difference in frequency and 
Hamiltonian, and the profile shows broader exponential tails. 
The same happens for the odd tip mode  Fig.\ \ref{fig:kerrloc} (f), which is 
also 
not directly connected to the compact mode. Finally, the family of 
odd line modes possesses {\em one compact mode} 
for the specific parameter relation $(\Gamma,\bar{\lambda}_c,J)$, 
which is shown in 
Fig.\ \ref{fig:kerrloc} (g). Thus, the nonlinear compact modes appear as 
special cases of the continuous family of stable, 
exponentially localized, modes centered on a line site, which can be tuned to 
compactification at these specific parameter values. 

For larger $J$, the odd line mode has an interval of instability around 
$2.43 \lesssim J \lesssim 5$, where it changes its form (upper bifurcation not shown in Fig. \ref{fig:kerrloc}(c)). 
The tip sites next to the central line site grow and pass its amplitude. 
A similar instability interval appears for the family of odd tip modes 
for smaller $J$,  $1.7 \lesssim J \lesssim 2.05$. In the interval between the 
two 
bifurcation points where the odd tip modes stabilize and the odd line 
modes destabilize, a family of intermediate, asymmetric modes appears 
(cf.\ Fig.~\ref{fig:kerrloc}(h)). This unstable mode, shown in 
Fig.~\ref{fig:kerrloc} with a dotted line in its 
regime of existence, is responsible for 
transferring the instability between the 
odd tip and line modes, and merges with the 
odd modes at the bifurcation points.
In the inset of Fig.~\ref{fig:kerrloc}(d) we show the difference between 
$H$ for the odd line and 
tip mode, where the crossing hints at this stability exchange  
scenario and the existence of the intermediate solution. Note however, that 
this regime is rather far from the compactification regime.

\begin{figure}[t]
\includegraphics[width=0.5\textwidth]{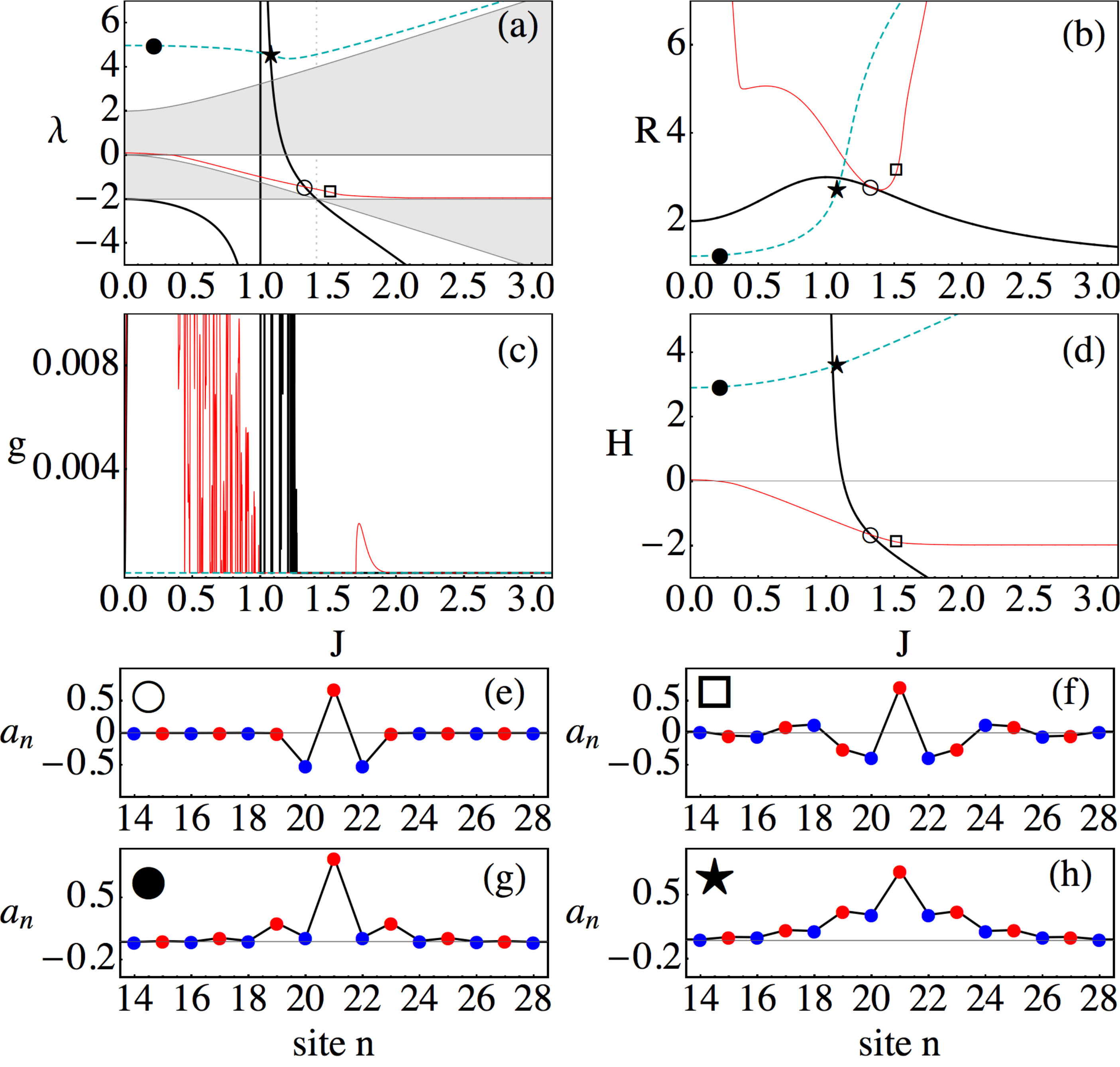}
\caption{(Color online) Properties of focusing nonlinear modes: 
(a) frequency $\lambda$, (b) $R$, (c) stability parameter $g$  and (d) $H$ 
vs.\ $J$ for the 
compact mode (thick black), the compactifying odd line gap mode when 
$\Gamma=1.2$ (thin red), and the non-compactifying odd line mode above the 
spectrum when $\Gamma=5$ (dashed green). Examples of stable 
modes are shown below: 
gap modes in (e)-(f) and modes above the spectrum in (g)-(h). 
}
\label{fig:kerrfoc}
\end{figure}
%
The analogous compactification of a nonlinear localized, stable 
{\em gap mode} for {\em focusing} nonlinearity is illustrated in 
Fig.\ \ref{fig:kerrfoc} (for simplicity, we include in this figure 
only modes with main peak at a single line site). 
It is important to stress that 
the compactifying mode, with its sign changing amplitudes 
(Fig.\ \ref{fig:kerrfoc} (e), (f)) is not 
identical to the fundamental odd line mode with amplitudes of the 
same sign (Fig.\ \ref{fig:kerrfoc} (g), (h)) and frequency above the spectrum, 
and they may both exist as simultaneously stable solutions. 
The condition to have the compactifying mode stable at the point where it 
compactifies requires the nonlinearity to be relatively weak, 
$\Gamma \lesssim 2.3$ (cf.\ (\ref{Pcubic})). On the other hand, the fundamental 
odd line mode exists as a stable solution also for larger $\Gamma$, as 
illustrated in Fig.\ \ref{fig:kerrfoc} for $\Gamma=5$, where the compact mode 
also has a frequency above the linear spectrum but is unstable. The properties 
of the compactifying modes for stronger focusing 
nonlinearities will be discussed 
further for the saturable nonlinearities in Sec.\ \ref{sec:saturable}
(cf.\ Fig. \ref{fig:compactsat1}).

\section{Saturable nonlinearity}
\label{sec:saturable}
Here, we consider a saturable nonlinearity, taken in equivalent form as in, 
e.g., Ref.\ \cite{NVS11} and references therein,
$$f(x)=  \frac{x}{1+x}\ .$$ Thus, in 
the small-amplitude limit, the Kerr nonlinearity of Sec.\ \ref{sec:cubic} is 
reproduced, while saturability effects become important for larger 
amplitudes. Note in particular that in the large-amplitude, strongly saturated, 
limit, $f(x) \rightarrow 1$, so the system (\ref{dnls}) 
becomes again effectively linear with just a frequency shift of the 
linear dispersion relation (\ref{eq:disp}), 
$\lambda \rightarrow \lambda + \gamma$. 

We then obtain for the left-hand side in the general compacton condition
(\ref{gencond}):
$$
\frac{\gamma}{J_l} 
\frac{|A|^2(1-1/|J|^2)}
{\left(1+|A|^2 \right) \left(1+|A/J|^2 \right)},
$$
which gives a second-degree equation for $|A|^2$. Solving this and using 
Eq.\ (\ref{power}) gives the following expression for the total compacton 
power, as function of the coupling ratio $J$ and the rescaled
nonlinearity/saturability parameter $\Gamma_s\equiv \gamma/J_l$: 
\begin{eqnarray}
P_{\pm} &&= \frac{1}{2}\left(1+2 / \left |J\right|^2\right)
\left[\Gamma_s 
\frac{\left |J\right |^2-1}
{2 -J^2}-1 - \left |J\right|^2
 \right.\nonumber\\&&\pm\left.\sqrt{\left(\Gamma_s 
\frac{\left |J\right |^2-1}
{2-J^2} -1 - 
\left |J\right|^2\right)^2-4 \left|J\right|^2 }
\ \right].\quad \label{Psatur}
\end{eqnarray}
It is important to stress that $\Gamma_s$ and $P$ 
act as two independent parameters for the saturable case, in contrast to the pure 
power-law case from Sec.\ \ref{sec:power-law} 
which only depends on the combination $(\gamma/J_l)^{1/\sigma}P$. 
Thus, a saturable nonlinearity introduces an additional, qualitatively 
different, mechanism for compactification tuning via the saturability 
parameter $\Gamma_s$.

We can note some general features of the expression (\ref{Psatur}): 
for given generic parameter values, there
are either zero or two possible solutions with different power (but note of 
course that only solutions with $P>0$ have physical meaning as compacton 
solutions). Also, for real $J$, the low power branch $P_-$ and the high power 
branch $P_+$ are positive 
within the same parameter space. Thus, an important difference to the cubic 
case is that the saturable nonlinearity promotes the appearance of 
{\em two 
different compact modes for a given value of $J$}, for the same existence 
regions. 
These two solutions possess the same profile (equal $\alpha$), but different 
amplitude $A$.

\begin{figure}
\includegraphics[width=0.5\textwidth,angle=0]{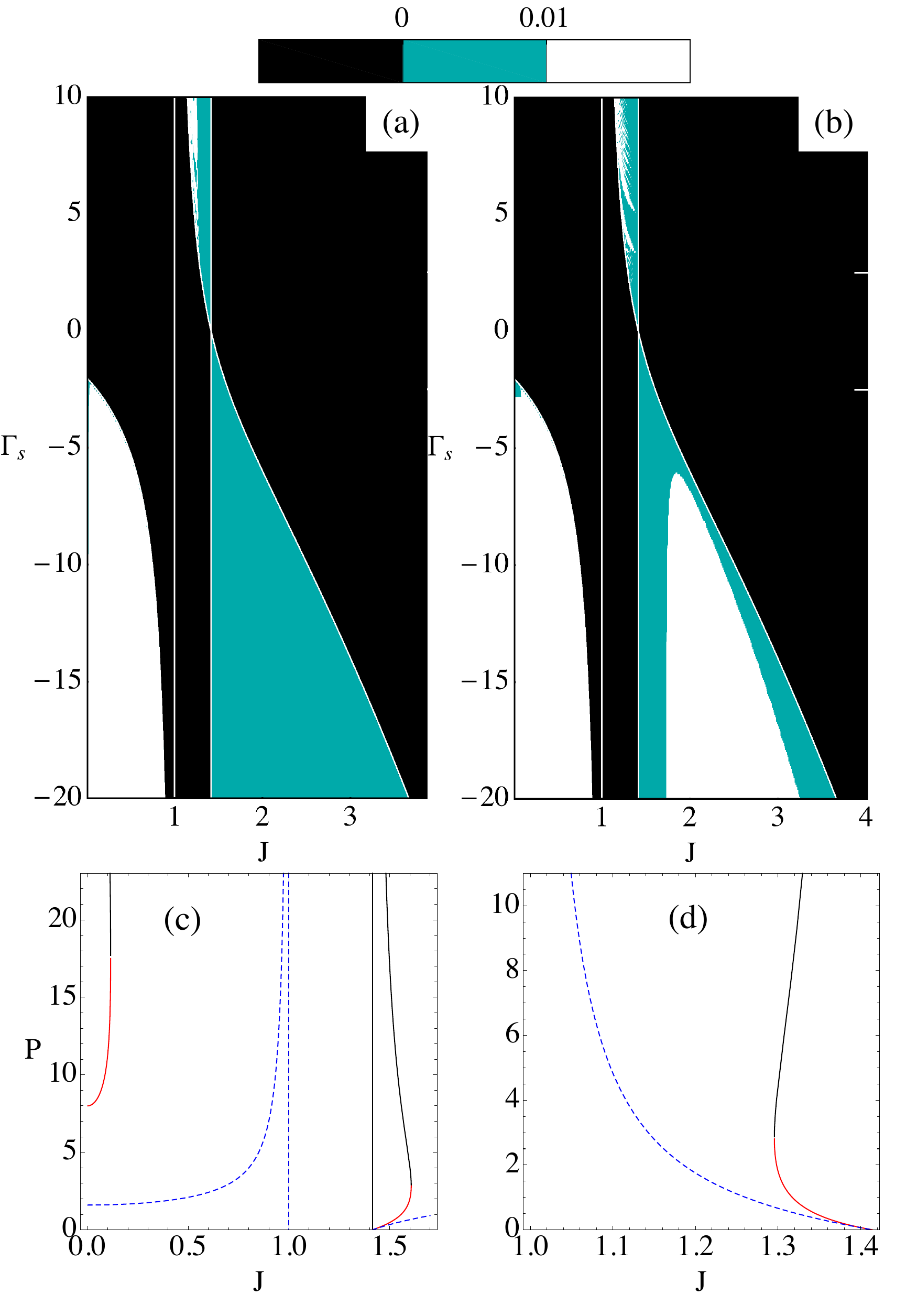}
\caption{(Color online) Top (a) and (b): white and shaded (green) areas show the existence 
regimes for exact compactons with saturable nonlinearity where $P>0$ from 
(\ref{Psatur}) (black regions imply no solutions). The non-vertical boundaries 
(white lines) show $\gamma_{bif}^+$ from (\ref{gammabif}). Additionally, we 
indicate the stability $g$ for (a) $P_-$ and (b) $P_+$ solutions, where 
stable solutions are contained in the shaded (green) regions ($g<0.01$), while 
white regions are unstable ($g>0.01$).
Bottom: two examples of power vs.\ coupling ratio $J$ for $\Gamma_s= -2.5$ 
(c) and  $\Gamma_s= 2.5$ (d). Solid lines show the $P_-$ (lower red) 
and $P_+$ (upper black) branches, while dashed blue line shows the 
corresponding expression for the cubic case (\ref{Pcubic}).
}
\label{fig:compactsat}
\end{figure}

Bifurcation points,  where two solutions coincide, appear when 
the expression under the square root in (\ref{Psatur}) is zero, which happens 
when 
\begin{equation}
\gamma_{bif}^{\pm}  \equiv \Gamma_s = \frac{\left( J^2-2\right) 
\left (1 \pm \left|J\right|\right)}
{(1 \mp |J|)}\ .
\label{gammabif}
\end{equation}
For a given  $\Gamma_s$, 
this gives a third-degree equation for the coupling 
ratios $J$ where bifurcations appear. Note, in particular, that when 
$\Gamma_s= - 2 $, one bifurcation point 
appears at $J=0$ (with $P\rightarrow\infty$). 
The consequence is that, in addition
to the two branches of solutions continuing the linear solution at 
$J = \sqrt{2}$, 
two new branches appear for small $J$ when $\Gamma_s< - 2$ 
(see Fig.\ \ref{fig:compactsat} (c), and lower left parts of (a) and (b)). 
Note also that at bifurcation points, the power is given by 
$$
P_{\pm} = \left|J\right|\left(1 + 2/ \left|J\right|^2\right)
,
$$
and $|A|^2 = |J|$. 

\begin{figure}
\includegraphics[width=0.5\textwidth,angle=0]{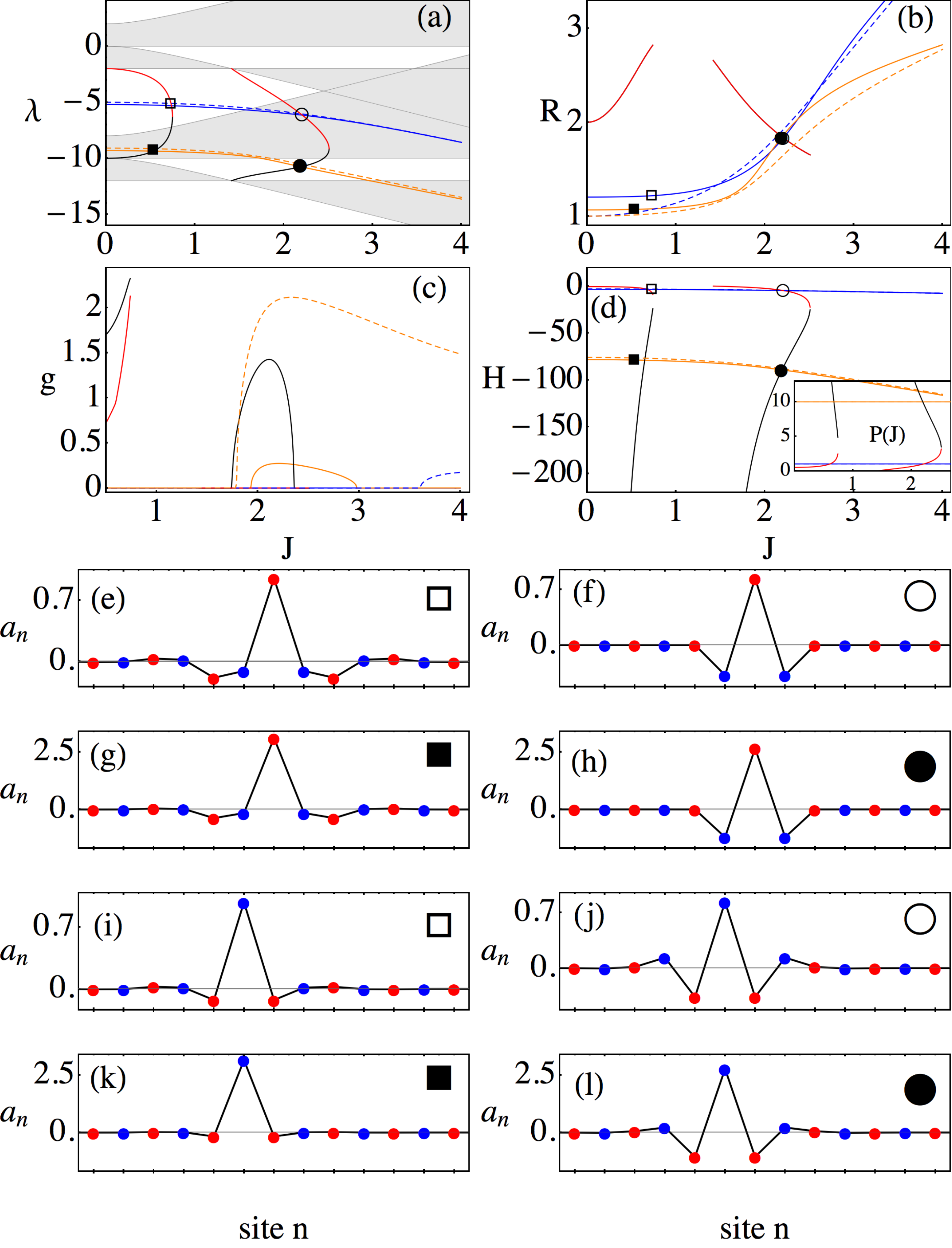}
\caption{(Color online) Diagrams for saturable nonlinear solutions with defocusing 
nonlinearity $\Gamma_s=-10$. (a) $\lambda$ versus $J$,
(b) $R$ versus $J$ (c) $g$ versus $J$, and (d) $H$ versus $J$, with inset 
 $P$ versus $J$. 
Exact compactons are shown in Red ($P_-$) and Black ($P_+$) lines, while odd line localized solutions are shown in Blue and Orange for $P=1$ and $P=10$, respectively. Dashed lines correspond to odd tip solutions for same parameters. Profiles for different powers and $J$ (marked by empty and filled squares and circles) are shown for (e)--(h) odd line and (i)--(l) odd tip modes. 
In (a) the linear ($P\rightarrow 0$, upper two bands) and 
saturated ($P\rightarrow \infty$, lower two bands) linear band spectra are 
shown with shaded areas. 
}
\label{fig:compactsat4}
\end{figure}

\begin{figure}
\includegraphics[width=0.5\textwidth,angle=0]{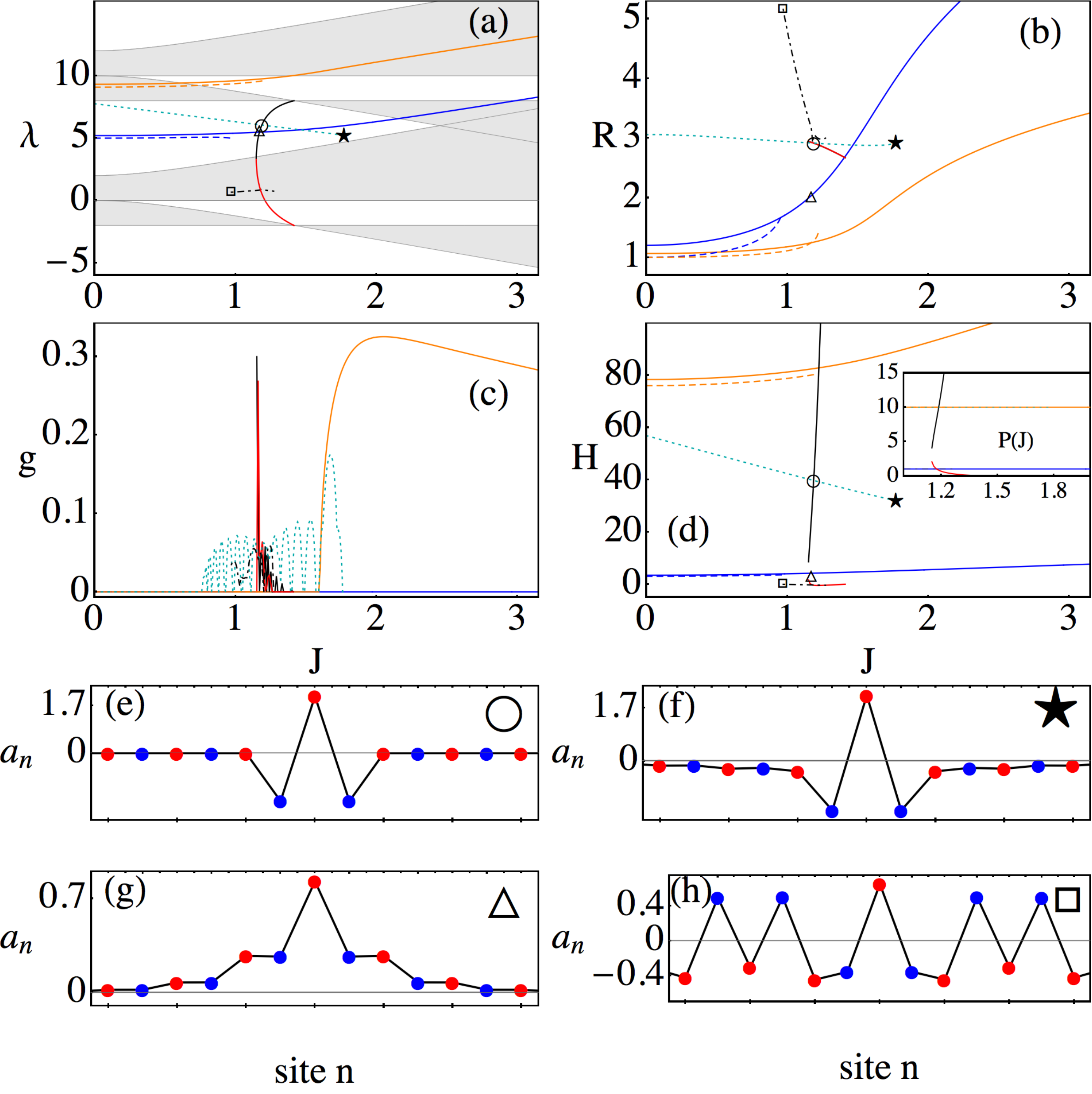}
\caption{(Color online) Diagrams for saturable nonlinear solutions with focusing 
nonlinearity, $\Gamma_s=10$.
(a)-(d) show the same quantitites as Fig.\ \ref{fig:compactsat4} (a)-(d) 
with the same line coding for solutions. 
In addition, 
compactifying odd 
line localized solutions are shown in black dotdashed for $P=1$ and 
green dotted for $P=10$. 
Profiles at positions indicated by markers: 
(e), (f)  modes from the compactifying family at $P=10$ 
(localized modes above linear spectrum); 
(g) the fundamental 
(noncompactifying) stable odd line mode at $P=1$;
(h) mode from the compactifying family at $P=1$ (non-localized 
mode inside upper linear band).
Shaded areas in (a) represent dispersion bands in the linear 
($P\rightarrow 0$, lower two bands) 
and saturated ($P\rightarrow \infty$, upper two bands) limits. 
}
\label{fig:compactsat1}
\end{figure}

%
From Fig.\ \ref{fig:compactsat} we can see the limits of existence 
(non-black regions in top figures) 
of the compacton solutions vs.\ $J$ and $\Gamma_s$. 
For $\Gamma_s\le - 2$, there are two windows of existence delimited by 
$\gamma_{bif}^+$ (white line) and $J_{lin}=\sqrt{2}$. For $\Gamma_s>- 2$, there 
exists only one region of compactons delimited by $\gamma_{bif}^+$  and 
$J_{lin}=\sqrt{2}$. 
Two examples for the saturable compacton power (\ref{Psatur}) as a function of 
$J$ for $\Gamma_s=- 2.5$ (c) and $\Gamma_s= 2.5$ (d) are shown on the bottom of Fig.\ \ref{fig:compactsat}, together with the corresponding expression for the cubic case (\ref{Pcubic}), 
which only agrees with the saturable case in the small-power regimes, as it 
should. The region in parameter space of $J$, where the lower saturable compacton branch $P_-$ is 
well approximated by the cubic compactons, increases with increasing 
$|\Gamma_s|$, 
as exemplified for $\Gamma_s = \pm 10$ in the insets of 
Figs.\ \ref{fig:compactsat4}-\ref{fig:compactsat1} (d).
Note however that, in contrast to the cubic case, 
for each fixed $\Gamma_s$ 
there is always an upper limit for $|J|$ for compacton existence along the 
branch with  $|J| > \sqrt{2}$ bifurcating from the linear compacton when 
$\Gamma_s < 0$. This limit, given  by 
$|J_{max}|=J(\gamma_{bif}^+)$, 
grows monotonically with $|\Gamma_s|$, and it follows from (\ref{gammabif}) that $|J_{max}| \sim \sqrt{-\Gamma_s}$ for large negative $\Gamma_s$.
Note also that, as $\Gamma_s$ decreases towards large  
negative values,  the additional branch for small $J$ born 
at $\Gamma_s=-2$ approaches the branch with $|J|<1$ appearing for 
$\Gamma < 0$ in the cubic case (Sec.\ \ref{sec:cubic}). However, 
in contrast to the cubic case also this branch has an upper limit 
for $|J| < 1$, and it can be seen  from (\ref{gammabif})
that $|J_{max}| \sim 1 - 2/ |\Gamma_s|$ 
for large negative $\Gamma_s$.
When $\Gamma_s > 0$, the condition to have 
$P>0$ demands that compactons may only appear for  
$|J| < \sqrt{2}$, just as for 
the cubic case in Sec.\ \ref{sec:cubic}. However, as seen 
in Fig.\ \ref{fig:compactsat}, this solution branch now has a 
lower limit with $|J| > 1$, and from   (\ref{gammabif}) we now obtain 
$|J_{min}| \sim 1 + 2 /\Gamma_s$ 
for large positive  $\Gamma_s$. 

Analogously to the studies for the cubic nonlinearity illustrated in 
Fig.\ \ref{fig:kerrcomp},
we have analyzed numerically the 
stability of the two saturable compacton solutions. The results are
summarized in Fig.\ \ref{fig:compactsat} (explicit examples for 
$\Gamma_s = \pm 10$ are shown in 
Figs.\ \ref{fig:compactsat4}-\ref{fig:compactsat1} (c)). When $\Gamma_s<0$ 
(effectively defocusing nonlinearity) and $J>\sqrt{2}$, 
the low power ($P_-$) branch is, as in the defocusing 
cubic case, always stable, independent of 
the $\Gamma_s$-value. Decreasing $\Gamma_s < 0$, the high-power ($P_+$) 
solutions destabilize for 
intermediate values of $J$, but generally regimes of stable compactons remain 
for low and high $J$ 
(the former corresponding to higher power and stronger saturation).
Thus, for defocusing saturable nonlinearities {\em two different stable 
compacton modes with different powers}, having similar profiles 
(equal value of $R$)  may exist for a given $\Gamma_s$.
Fig.\ \ref{fig:compactsat} also shows that both solutions appearing for 
$J<1$ when $\Gamma_s < -2$ 
always are unstable.
Similarly to the cubic case, this regime is not connected with flat-band 
linear modes.
When $\Gamma_s>0$ 
(effectively focusing nonlinearity), a similar scenario as in the focusing 
cubic case is observed for
the compact mode continued from the linear limit at 
$J=\sqrt{2}$: it is stable for small powers but destabilizes through 
resonances with extended lattice modes for larger powers. However, similarly 
to the defocusing saturable case, the compact mode generally restabilizes in 
the strongly saturated, high-power regime of the $P_+$ branch, and thus we may 
also for the focusing case find regimes of two simultaneously stable 
compactons, although only in a narrow interval for $J \lesssim \sqrt{2}$.

To study the relation of these compact solutions with families of standard, 
non-compact, localized modes (cf.\ Fig.\ \ref{fig:kerrloc} for the cubic 
case), we first present in Fig.~\ref{fig:compactsat4} collected results 
for $\Gamma_s=-10$, i.e., an effectively defocusing case with 
two different regions for analytical solutions.
First, as discussed above, compact solutions are seen 
to bifurcate at $J=\sqrt{2}$ from linear flat band modes with zero ($P_-$) and 
infinite ($P_+$) power, coinciding with the low and high amplitude (saturated) 
linear spectra 
(the latter being shifted to $\lambda+\gamma$ due to saturation) as 
shown in Fig.~\ref{fig:compactsat4}(a). 
For $J > \sqrt{2}$, we see that both $P_-$ and $P_+$ compacton branches 
(but not the branches with $J<1$) appear 
as compactifications of families of standard localized modes with main 
localization at the central line site. 
Analogously to the cubic case in Sec.\ \ref{sec:cubic}, we illustrate in 
Fig.~\ref{fig:compactsat4} properties of numerically obtained families 
of modes centered at line and tip sites, respectively, at two different fixed 
values 
of power: $P=1$ and $P=10$.  
Figs.~\ref{fig:compactsat4}(e)--(l) show profiles for low power solutions (empty squares and empty circles) and high power ones (black squares and black circles), for different values of $J$ as well as for line and tip site centers. 
Note that, while the low-power compactifying line-site centered mode 
(Figs.~\ref{fig:compactsat4}(e)--(f)) is always 
stable for the relevant regime in $J$, just as in the cubic case, the 
corresponding family of high-power modes for $P=10$ 
(Figs.~\ref{fig:compactsat4}(g)--(h)) has an instability window around 
the compactification point, and thus this compact mode is unstable. This 
instability window moves as the power is further increased, implying that, 
as mentioned above, also the high-power compacton branch stabilizes for 
a sufficiently high power (for $\Gamma_s=-10$, high-power compacton
restabilization  appears for $P \gtrsim 25.36$, corresponding to 
$J \lesssim 1.75$). 
Note that also the non-compact modes with exponential tails 
are strongly localized (small $R$) 
for these parameter values, as they are residing far below the linear band.
%
%

In Fig.\ \ref{fig:compactsat1}, we illustrate in an analogous way the case 
with focusing 
nonlinearity, $\Gamma_s=+10$. 
As in the cubic case, neither the upper nor lower compacton branches
then appear as compactifications of the family of fundamental localized modes
with constant  sign and main localization at the central line site, 
which is seen to appear simultaneously as a stable solution 
but with different properties 
(Fig.\ \ref{fig:compactsat1} (g)). For small powers, when the compacton 
frequency is still in the linear gap, the scenario is similar to the cubic 
case illustrated in Fig.\ \ref{fig:kerrfoc} with a compactifying gap mode with 
a tail of oscillating signs (cf.\ Fig. \ref{fig:kerrfoc} (f)). For slightly 
larger powers ($P=1$ in Fig.\ \ref{fig:compactsat1}), when the compacton 
frequency has entered the upper linear band but still belongs to the $P_-$ 
compacton branch, the compactifying family is not localized but has a 
non-decaying, oscillating tail (Fig.\ \ref{fig:compactsat1} (h)), 
which vanishes only at the compactification point. As seen in 
Fig.~\ref{fig:compactsat1} (c), these solutions are generally unstable. 
Finally, when the power is large enough for the compacton 
frequency to reside above the linear spectrum (belonging then to the $P_+$ 
branch for the case in Fig.\ \ref{fig:compactsat1} (a)), the compactifying 
family is again a localized solution with main localization at three sites, 
where only the central site has a different amplitude sign than the other 
sites (Fig.\ \ref{fig:compactsat1} (f)). From Fig.\ \ref{fig:compactsat1} (c), 
this solution is seen to be stable in parts of its existence regime 
(e.g., the mode in (f)), but for these parameter values, it is unstable 
at the point where it compactifies. However, for even larger $P$, at strong 
saturation, these solutions will also be stable at the compactification point, 
as illustrated in Fig.\ \ref{fig:compactsat} (b). This 
{\em restabilization of the compactifying focusing mode above the linear 
spectrum 
at high power} is another 
characteristic feature of the saturable system, which does 
not appear for the cubic case.

\section{Further tuning options}
\label{sec:further}

Let us now consider a generalization of Eq.\ (\ref{dnls}), by adding
a non-homogeneous 
distribution $\Delta_n$ of on-site energies, and allowing also for 
anisotropies in the coupling coefficients
$J_{n,m}$,  
\begin{equation}
i \dot{u}_n + \Delta_n u_n +\sum_m J_{n,m} u_m + \gamma f(|u_n|^2) u_n =0.
\label{anisodnls}
\end{equation}
We were able to find three-site compact modes, having the same symmetric 
line-site 
centered structure as described in Sec.\ \ref{sec:general},  
for a distribution of 
\begin{equation}
\Delta_n/J_l=\begin{cases}0 & \text{ for $n$ odd}\\ \Delta& \text{ for $n$ even.}\end{cases}
\label{eq:Delta}
\end{equation}
This means, that the tips have an on-site energy of $\Delta$, whereas the line 
sites have vanishing on-site energy (the linear dispersion relation and the 
corresponding linear compact mode in the isotropic sawtooth lattice 
with $\Delta = -1$
were illustrated in Ref.\ \cite{Flach14}, Fig.\ 1 (f)). 

\begin{figure}[t]
\includegraphics[width=0.5\textwidth]{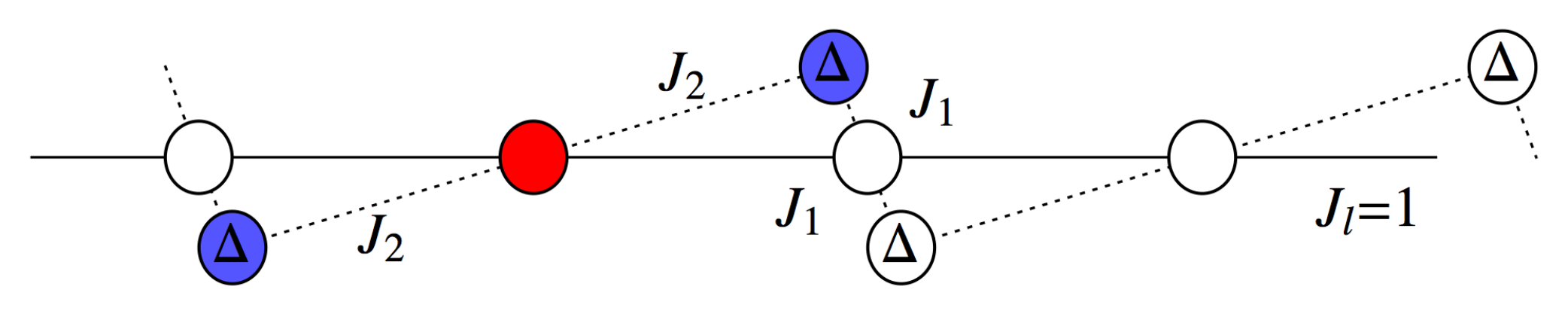}\\
\caption{(Color online) Sketch of the anisotropic geometry with pairwise 
alternating couplings and alternating on-site energies. 
}
\label{fig:anisch}
\end{figure}
Considering also various ways 
to introduce coupling anisotropies in the sawtooth lattice, one possibility 
would be to have constant line coupling with alternating couplings to the tips. 
However, such a structure only allows for asymmetric linear compactons 
for some specific ratio of the two tip coupling constants, and no compact 
nonlinear continuations were found for these modes for any distribution of 
on-site 
energies of the type (\ref{eq:Delta}) with identical tip energies 
(nonlinear asymmetric compactons may exist under special conditions 
if the alternating coupling is also 
combined with different on-site energies for up- and down- pointing tip sites,
but these conditions are more complex and will not be discussed further here). 
On the other hand, a nonlinear continuation of symmetric compact modes is  
possible, if instead a pairwise (real) alternating coupling $J_1$ and $J_2$ is 
assumed in  the tips, as sketched in Fig.\ \ref{fig:anisch}.  
Within the line we conserve the coupling $J_l=1$.  In this case, the four linear bands are given by  
\begin{eqnarray}
&&\lambda_{\pm,\pm}=\frac{\Delta}{2} + \cos (2 k)\pm\label{dispanisoj}\\
&& \sqrt{ (J_1^2+J_2^2)+\left(\frac{\Delta}{2} - \cos (2 k)\right)^2\pm f_{J_1,J_2}(k)}\nonumber\\
&&\nonumber\\
&&\text{with}\nonumber\\
&&\nonumber\\
&&f_{J_1,J_2}(k)=\sqrt{J_1^4+J_2^4-2J_1^2 J_2^2+ 4J_1^2 J_2^2 \cos^2(2 k)}\nonumber
\end{eqnarray}

The geometry allows for two different types of compact modes, 
one that has the line site coupled to the tips with $J_1$, 
the other with $J_2$. 
Without loss of generality we will consider only the latter, shown 
exemplary in Fig.~\ref{fig:anisch}. 
For a Kerr nonlinearity, the symmetric compact modes  
with amplitude ratio $\alpha=-1/J_1$ appear when the following conditions
for the effective nonlinearity (\ref{Pcubic}) and frequency are fulfilled,
\begin{eqnarray}
\Gamma_{\Delta}(J_1,J_2) &=&\frac{J_2(J_1^4-4)-J_1\Delta(J_1^2+2)}{J_1-J_1^3}\nonumber\\&&\label{compdeltaj1j2}\\
\lambda_{\Delta}&=&\frac{\Delta  J_1^3+2J_2-J_1^4J_2}{J_1^3-J_1}\nonumber
\end{eqnarray}

\subsection{Alternating on-site energies}
\label{sec:onsite}

\begin{figure}[t]
\includegraphics[width=0.5\textwidth]{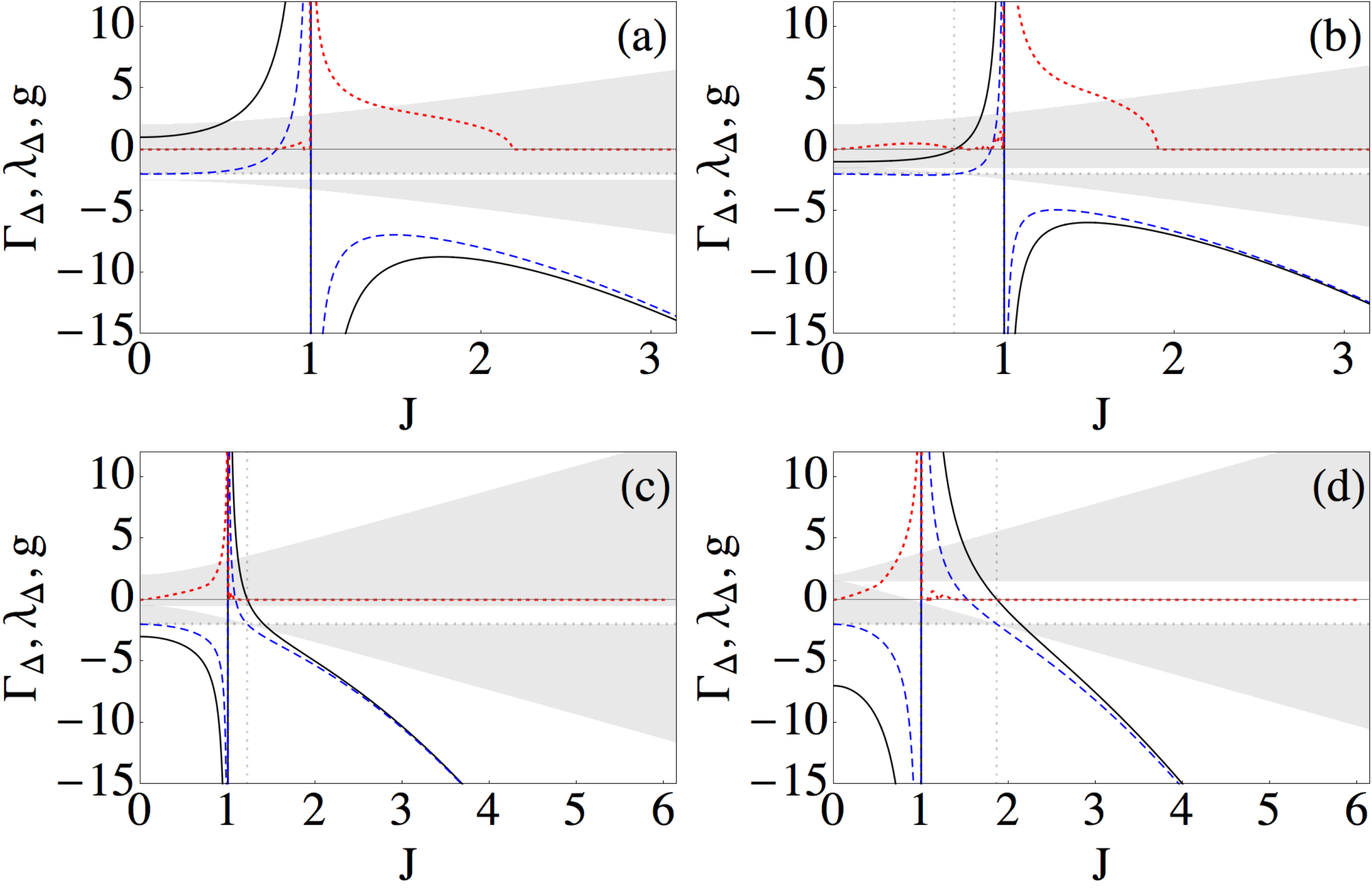}
\caption{(Color online) Compact mode $\Gamma_{\Delta}$ (solid), $\lambda_{\Delta}$ (dashed), 
and stability parameter $g$ (dotted) 
vs.\ $J$ for $\Delta=-2.5$ (-1.5,-0.5, 1.5) in (a)-(d). 
The shaded areas denote the band extensions. }
\label{fig:aniso}
\end{figure}
We first turn to the isotropic case of $J_1=J_2=J$, but $\Delta\neq 0$. 
The linear dispersion relation \eqref{dispanisoj} for this case is
\begin{eqnarray}
\lambda_\pm(k)&=&\frac{\Delta}{2}+\cos(2k) \pm\label{eq:dispaniso}\\ &&\sqrt{\left(\cos(2k)-\frac{\Delta}{2}\right)^2+2J^2[1+\cos(2k)]},\nonumber
\end{eqnarray}
whereas the Eqs.\ \eqref{compdeltaj1j2} simplify to
\begin{eqnarray}
\Gamma_{\Delta} &=&\frac{\left(J^2+2\right) \left(J^2-\Delta -2\right)}{1-J^2}\nonumber\\&&\label{compdelta}\\
\lambda_{\Delta}&=&\frac{\Delta  J^2+2-J^4}{J^2-1}.\nonumber
\end{eqnarray}
This gives a linear compact mode for $J_{\Delta,lin}=\sqrt{2+\Delta}$, thus 
there is only a continuation of the linear compact solution for 
$\Delta\ge -2$. The bandgap vanishes as well for $\Delta=-2$. 
However, nonlinear compact solutions exist also for $\Delta < -2$. 
In Fig.\ \ref{fig:aniso} we show different scenarios. For $\Delta=-2.5$ 
(see Fig.\ \ref{fig:aniso}(a)), there is no flat band and the compact mode 
does not cross the bandgap. For $\Delta=-1.5$ (see Fig.\ \ref{fig:aniso}(b)), the flat band is located for $J<1$, where the compacton frequency curve enters and crosses the bandgap with positive $\Gamma_{\Delta}$. The sign of the discontinuity at $J=1$ changes at $\Delta=-1$ and with it the location of the flat band, so for $\Delta=-0.5$ and  $\Delta=1.5$ (Fig.\ \ref{fig:aniso}(c)-(d))
the crossing of the bandgap is found for $J>1$. Increasing $\Delta$ further only opens the gap more, but there will be no further changes in regarding the bandgap.  Therefore, the on-site energy difference $\Delta$ gives rise to the possibility of a direct engineering of the bandgap as well as the nonlinearity and frequency of the compacton. 


\begin{figure}[t]
\includegraphics[width=0.5\textwidth]{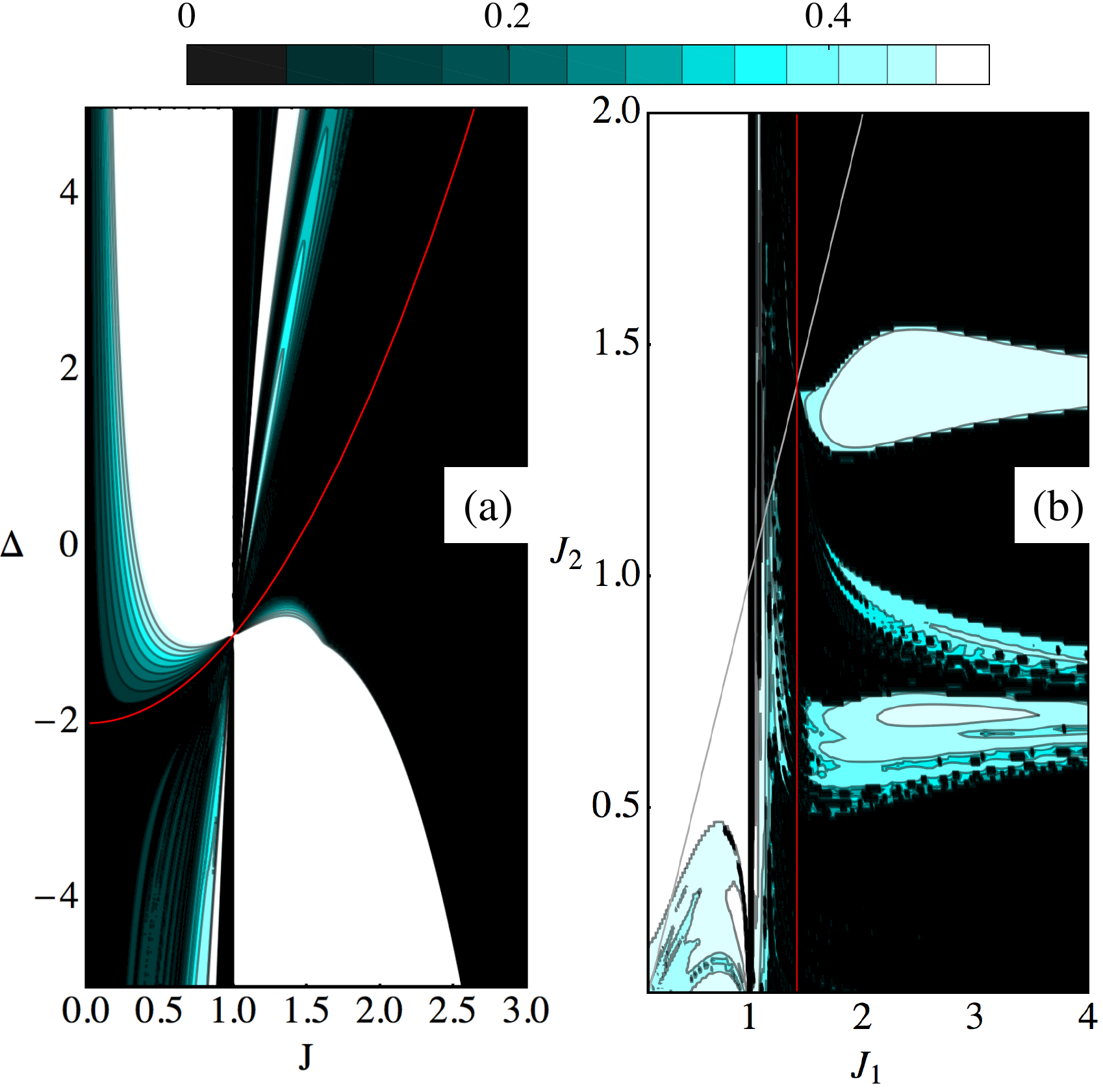}
\caption{(Color online) Stability parameter $g$ of the compact mode 
vs.\ $J_1=J_2=J$ and $\Delta$ (a), 
vs.\ $J_1$ and  $J_2$ with $\Delta=0$ (b), 
only the black areas ($g<0.05)$ contain stable modes. $N=40$. In (a) 
compactons are unstable  also in the black stripe for 
small $J$ when $\Delta>-2$, but have $g<0.05$.  
The linear compacton condition is given by red lines, the ordinary sawtooth with homogeneous  $J_2=J_1$  is marked with a gray line. 
}
\label{fig:anisostab}
\end{figure}
%
The stability for compact modes vs.\ $J$ and $\Delta$ are shown with dotted 
lines for the specific examples in Figs.\ \ref{fig:aniso}(a)-(d). 
The general stability scenario is explored in
Fig.\ \ref{fig:anisostab}(a), where black areas denote stable or weakly 
unstable ($g<0.05$) compactons and the 
degree of instability increases with brightness. As a guide to the eye, we 
denote the parameters of the linear compacton, $\Delta=\sqrt{J^2-2}$, with a 
red line. This line apparently separates two different regimes, 
with the main lobes of the instability located at opposite sides of $J=1$. 
For a fixed $\Delta<-2$ (Fig.\ \ref{fig:aniso} (a)), 
there is always a {\em stable regime for small $J$}, 
corresponding to 
a compacton with main amplitude at the {\em tip} sites in a chain with focusing 
nonlinearity. For growing $J$, some weaker instabilities are passed 
for $J<1$, and at $J=1$ the large instability lobe appears as the compacton 
now exists for defocusing nonlinearity.
By comparing with the stability results for the 
corresponding solutions in the open trimer configuration (without a 
surrounding lattice) \cite{Buonsante}, it can be deduced that 
the instabilities on the focusing side ($J<1$) are due to resonances with 
lattice oscillations (since this regime is always stable for the 
trimer \cite{Buonsante}), while on the defocusing 
side ($J>1$) it is an internal resonance in the compacton core which appears 
also for the trimer. 
In the regime $-2<\Delta<-1$ (Fig.\ \ref{fig:aniso} (b)), 
the main qualitative change appears in the 
small-$J$ regime, where the nonlinearity now is defocusing 
for $J<J_{\Delta,lin}$ and the compacton 
is unstable. In fact, if follows from analogous considerations for the open 
trimer \cite{Buonsante} that compactons with main localization on tip sites 
(small $J$) 
are always unstable through core instabilities for defocusing nonlinearities.
For $\Delta > -1$  (Fig.\ \ref{fig:aniso} (c)-(d)), 
the mode is always unstable in the defocusing regime $0<J<1$. When 
$\Delta \gtrsim -1$, we also see an instability in the other defocusing  
region for $J > J_{\Delta,lin}$, as a remains of the instability lobe at $J>1$ 
for smaller $\Delta$. For larger $\Delta$, these instabilities disappear 
and the only significant instabilities observed for  $J>1$ are those in 
the focusing regime $ J < J_{\Delta,lin}$, which are analogous to those described 
for the homogeneous case $\Delta=0$ in
Sec.\ \ref{sec:cubic}. In general, for all $\Delta$ the compacton 
always approaches a stable one-site line-mode for large $J$, corresponding 
to a strong defocusing Kerr nonlinearity. 

One important conclusion from the above analysis is, that by introducing 
alternating on-site 
energies with $\Delta < -2$ one may, for focusing nonlinearity, 
{\em stabililize also compactons with main localization at the tip sites}, 
which in the previous 
Sections were found to be always unstable for the homogeneous sawtooth lattice.

\subsection{Pairwise alternating couplings}
\label{sec:anisotropic}

We finally turn briefly to the case of pairwise alternating coupling as in 
Fig.\ \ref{fig:anisch}, but considering for simplicity 
only vanishing on-site energies 
$\Delta=0$. 
As the linear stability analysis vs.\ $J_1$ and $J_2$ for this family shows 
(see Fig.\ \ref{fig:anisostab}(b)), for small external coupling to the 
array $J_1\leq 1$ (corresponding to defocusing nonlinearity, $\Gamma < 0$
according to (\ref{compdeltaj1j2})) 
nearly all modes (having their main amplitudes on tip sites) 
are unstable, only a very restricted zone for small  $J_2 < J_1$ is stable. 
On the other hand, if $J_1>1$, there are as before 
many regimes of stable modes. In the focusing regime, $1 < J_1 < \sqrt{2}$ 
(see (\ref{compdeltaj1j2})), the stability scenario is qualitatively similar 
as for the case $J_2=J_1$ described in Sec.\ \ref{sec:cubic}; changing the 
internal coupling $J_2$ mainly changes the internal compacton oscillation 
frequencies 
and thereby also the exact location of the instability threshold resulting 
from their resonances with the surrounding chain, but a stable regime 
generally remains for small nonlinearity.  
In the defocusing regime $J_1 > \sqrt{2}$, the compacton having main 
amplitude on the line-site always remains 
stable for $J_2 \geq J_1$,  but it may destabilize in certain regimes for 
$J_2 < J_1$, i.e., when the internal coupling is weaker than the external. 
Note in particular the instability window around 
$J_2 = \sqrt{2}$, corresponding to a resonance with a linear compacton 
with internal coupling $J_1$.  

\section{Conclusions}
\label{sec:conclusions}

In this work we showed the existence of compact modes for generic realizations 
of the sawtooth lattice, considering power law and saturable nonlinearities. These modes are compact nonlinear continuations of the linear flat-band modes found in the sawtooth model. For a wide window in parameter space we showed the stability of these modes for both focusing and defocusing nonlinearities. Furthermore, we explored the influence of inhomogeneous on-site energies and pairwise alternating coupling on the band-structure and stability of the compact modes. We observed, that negative on-site energies or high anisotropic pairwise coupling increase the stability regimes. 

Moreover, we have confirmed the existence of compacton solutions for the 
saturable nonlinearity, finding stable regimes for two types of analytical 
solutions. As they are very compact solutions, they could be easily excited 
in an experiment, for low level of power and for different sawtooth geometries 
governed by the coupling ratio $J$. The possibility to excite 
very localized nonlinear solutions using low level of power is an important 
property of this particular lattice, in 
contrast with nonlinear conventional lattices~\cite{VJ13}.
We also found regimes with possibility to excite two, simultaneously stable, 
compactons with similar profiles, for small and large level of power. This is a very 
interesting property of the sawtooth lattice; in conventional systems, 
stable solutions for different level of power exist but with different profile structure (e.g., different participation number).

\begin{acknowledgments}
Authors want to thank A.J. Mart\'inez for fruitful discussions at the beginning of this work; U.N. thanks J. Calvo and D. Zueco for discussions of the linear properties. The research has been performed with support from the Swedish Research Council within the Swedish Research Links programme, 348-2013-6752.  U.N. appreciates the Spanish government projects FIS 2011-25167 and FPDI-2013-18422 as well as the Arag\'on project (Grupo FENOL). R.A.V. acknowledges support from Programa ICM grant RC130001, Programa de Financiamiento Basal de CONICYT (FB0824/2008), and FONDECYT Grant No. 1151444.

\end{acknowledgments}

\end{document}